\documentclass{PoS}
\usepackage{upgreek}
\usepackage{wrapfig}
\newcommand{\SNN}{$\sqrt{s_{_{{\rm NN}}}}=$}
\newcommand{\lbyl}{$\upgamma \upgamma \to \upgamma \upgamma$}
\newcommand{\ga}{\upgamma}
\newcommand{\ee}{\rm{e}^{+}\rm{e}^{-}}

\title{Measurement of Diffractive and Exclusive processes}

\ShortTitle{Diffractive and exclusive results by CMS}

\author{\speaker{Ruchi Chudasama (for the CMS Collaboration)}\\
        Indian Institute of Technology-Bombay and Bhabha Atomic Research Centre, Mumbai, India\\
        E-mail: \email{ruchi.chudasama@cern.ch}}

\abstract{With excellent performance the Compact Muon Solenoid (CMS) experiment has made a 
number of key observations in the diffractive and exclusive processes and hence in probing 
the Standard model in a unique way. This presentation will cover recent results on the 
measurement of diffractive and exclusive processes using data recorded by CMS detector at the LHC.}

\FullConference{The 39th International Conference on High Energy Physics (ICHEP2018)\\
		4-11 July, 2018\\
		Seoul, Korea}

\begin{document}

\section{Introduction}
The measurement of diffractive processes provide a valuable input for
the proper modeling of the full final state of hadronic interactions in event generators.
The exclusive processes provide a wide range of opportunities from testing 
QED, probing the gluon density inside the proton/nucleus to searches for physics beyond the Standard Model (SM). 
This report presents the latest results of diffractive and exclusive processes based on the 
data collected by the CMS experiment \cite{cmsdet}.
\section{Light-by-light scattering in PbPb collisions at \SNN 5.02 TeV}
Elastic light-by-light (LbL) scattering, \lbyl, is a pure 
quantum mechanical process that proceeds at leading order in the quantum electrodynamics (QED),
via virtual box diagrams containing charged fermions or boson. 
In the extension of the SM, it could also contain new charged super-symmetric particles, axions or monopoles. 
This report provides a study of the LbL process and new exclusion limits on axion-like particles (ALPs) production, 
using PbPb collision data recorded by the CMS experiment in 2015 at \SNN 5.02 TeV. 
The exclusive diphoton candidates are selected by requiring exactly two photons with E$_{\rm T}>$ 2 GeV 
and $|\eta| <2.4$. Further, events reconstructed with charged-particle tracks with p$_{\rm T}>$ 0.1 GeV 
and with calorimeter activity above noise thresholds are rejected. The non-exclusive diphoton background
is eliminated by selecting events with diphoton acoplanarity ${\mathrm A}_{\phi} < 0.01$
and diphoton transverse momentum p$_{\rm T}^{\ga \ga}< 1$GeV. 

\begin{wrapfigure}{r}{7.0cm}
\includegraphics[width=0.45\textwidth]{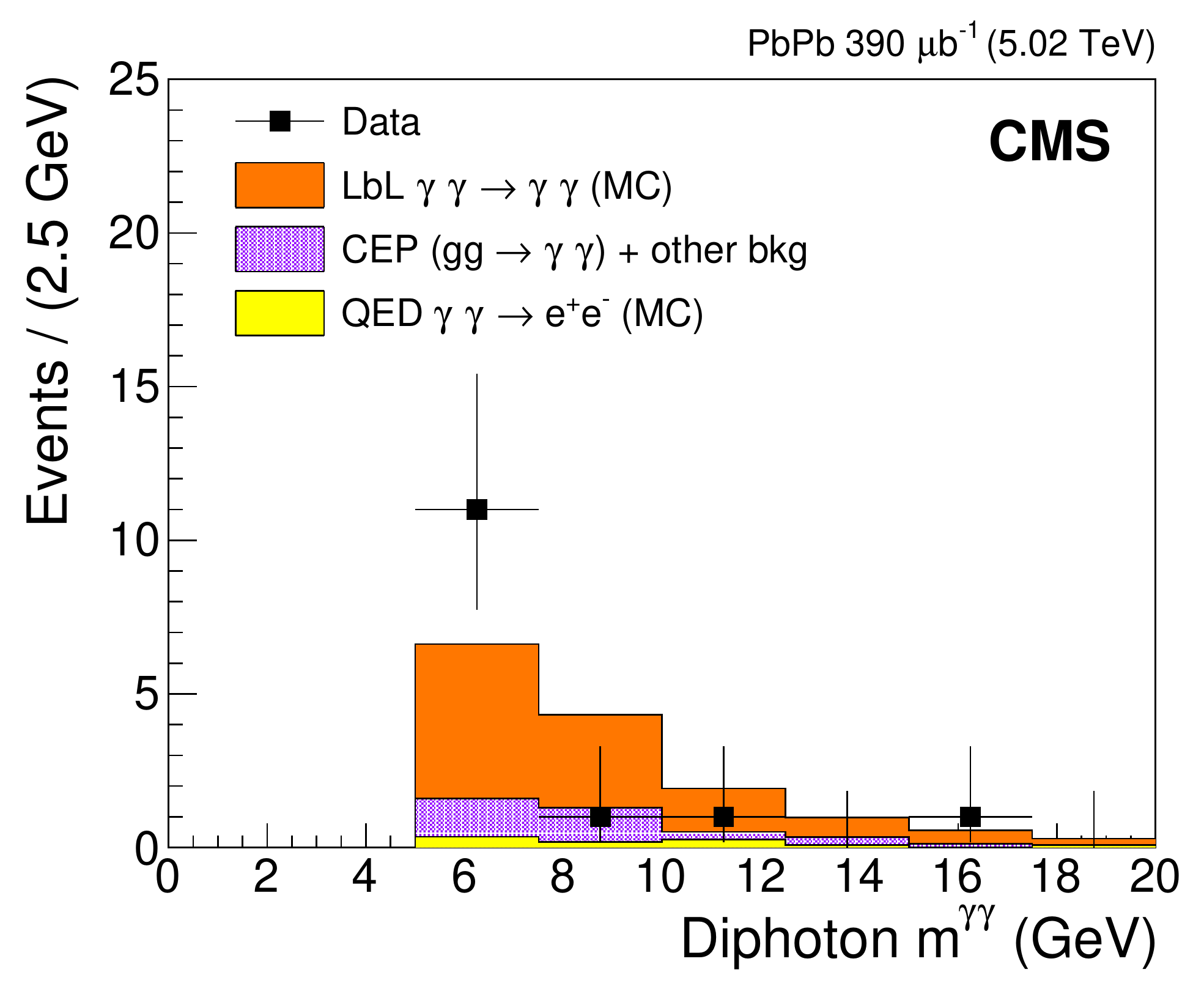}
\caption{Diphoton invariant mass for exclusive $\ga\ga$ events in data (squares) compared to MC expectations~\cite{lbyl_cms}.} 
\label{fig:control_plots}
\end{wrapfigure}

\begin{figure*}[hbtp]
\begin{center}
\includegraphics[width=0.40\textwidth]{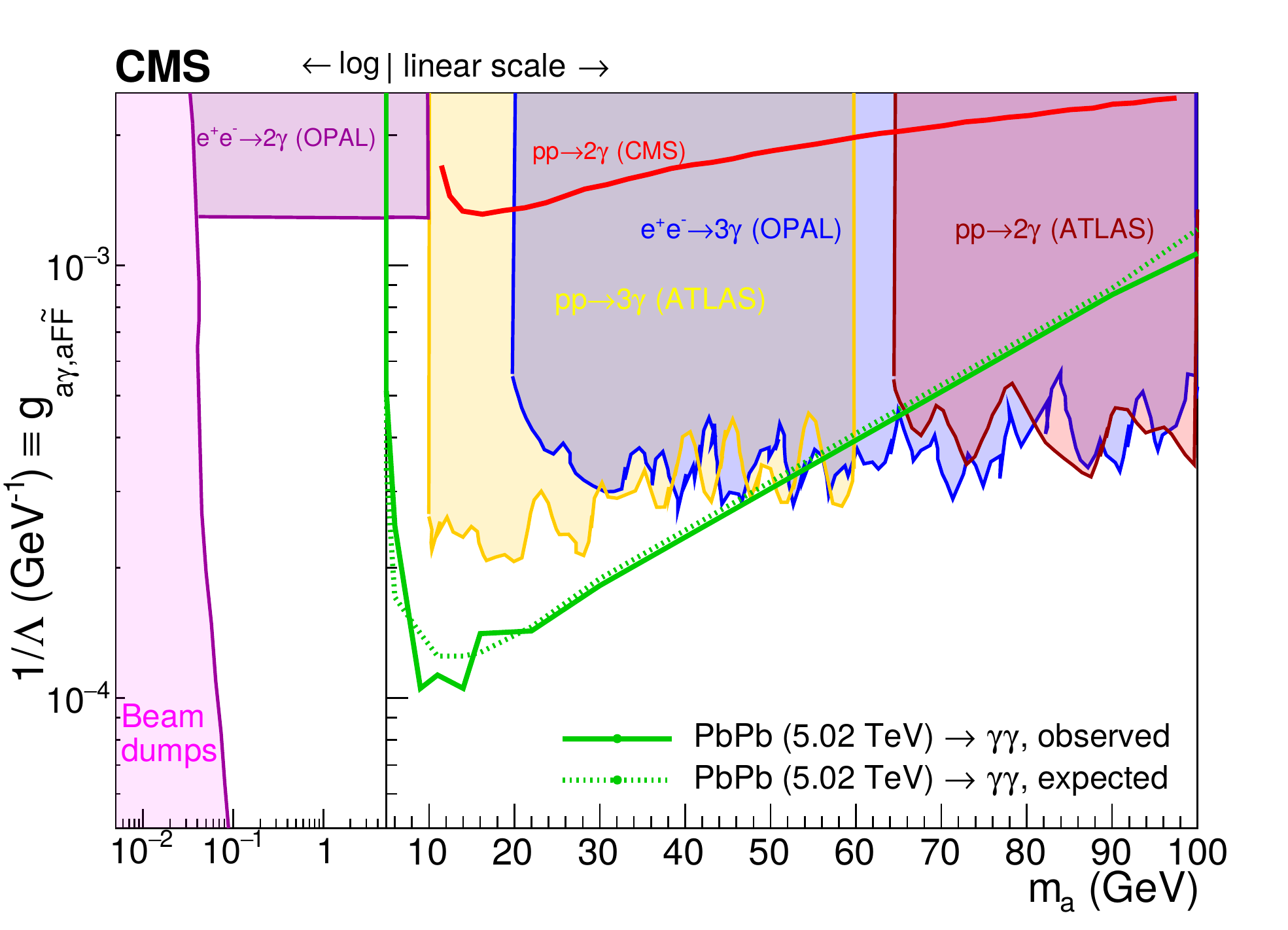}
\includegraphics[width=0.40\textwidth]{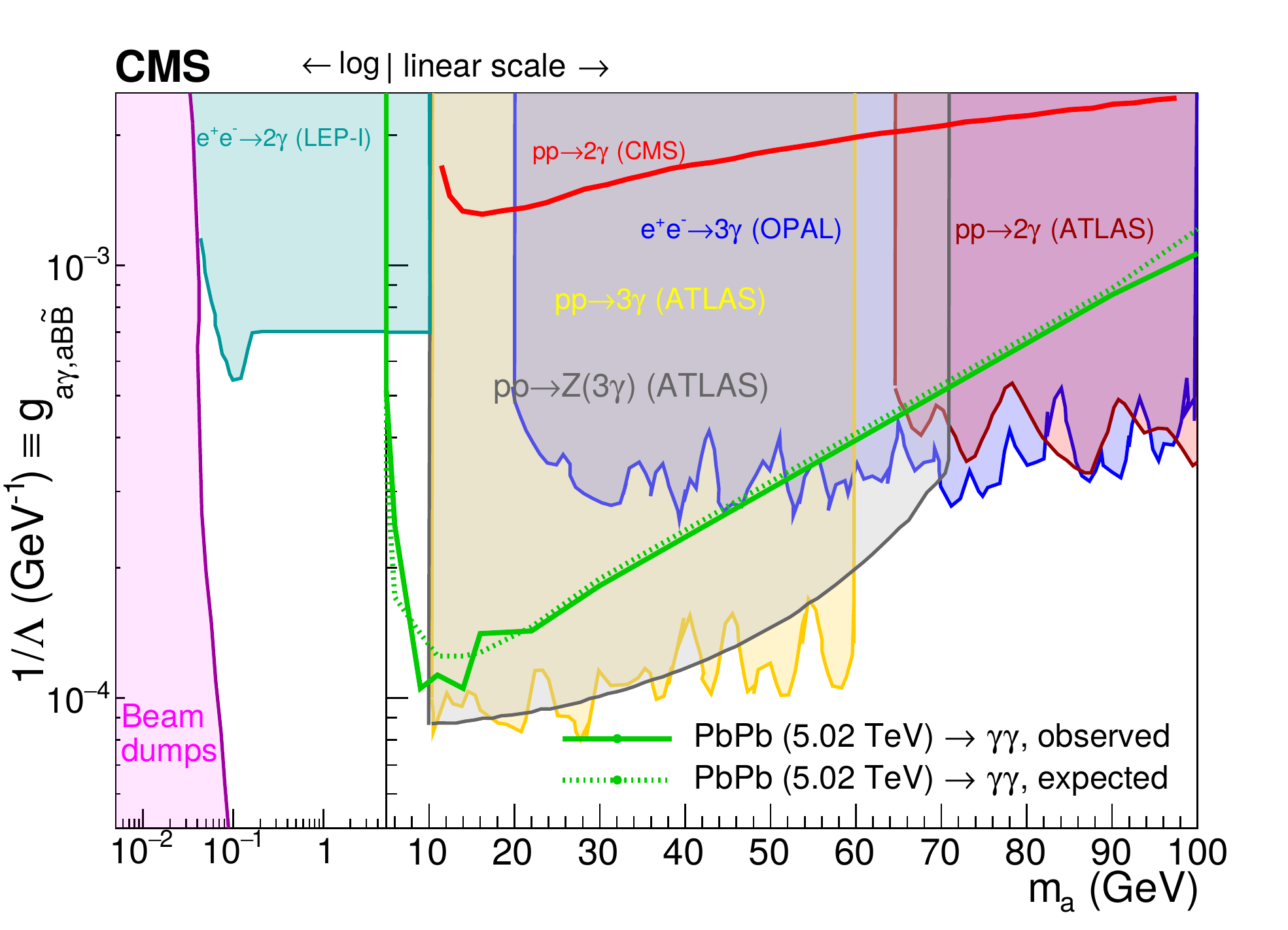}
\caption{Exclusion limits at 95$\%$ CL in coupling vs mass plane for (a) ALPs coupling to photons only 
(b) including also the hypercharge coupling ~\cite{lbyl_cms}.} 
\label{fig:axion_limits}
\end{center}
\end{figure*}

We observe 14 LbL scattering candidates, 
to be compared with 11.1 $\pm$ 1.1 (theo) expected from the LbL scattering signal 
(generated with {\sc Madgraph}), 3.0 $\pm$ 1.1 (stat) 
from CEP, and 1.0 $\pm$ 0.3  (stat) from QED $\ee$ background events. 
Fig.~\ref{fig:control_plots} shows the comparison of the measured and simulated 
diphoton invariant mass distribution, both the measured yields and kinematic distributions are in 
accord with the combination of the LbL signal plus QED $\ee$ and CEP+other background expectations. 
The observed (expected) signal significance is 4.1 (4.4) $\sigma$. 
The ratio $R$ of cross sections of the LbL scattering 
over the QED $\ee$ processes was measured, and amounts to $R = (25.0 \pm 9.6 \textrm{(stat)} \pm 5.8 \textrm{(syst)}) \times 10^{-6}$. 
The LbL fiducial cross section is obtained from the theoretical prediction of 
$\sigma(\ga\ga \to \ee, m^{\rm{e}\rm{e}}>5 \textrm{GeV}) = 4.82 \pm 0.15 \textrm{(theo)}$ mb and estimated to 
be $\sigma_\textrm{fid} (\ga\ga \to \ga\ga) = 120 \pm 46 \textrm{(stat)} \pm 28\textrm{(syst)}\pm4\textrm{(theo)}$ nb, 
in good agreement with the theoretical LbL prediction. 
The measured invariant mass distribution (Fig.\ref{fig:control_plots})  is used to search for pseudoscalar ALPs
produced in the process $\ga\ga\to\rm{a}\to\ga\ga$. 
The exclusion limits at 95$\%$ confidence level is applied 
on the cross section and used to set exclusion limits in the g$_{\mathrm{a}\upgamma}$ (coupling) vs, 
m$_{\rm{a}}$ (axion mass) plane. Fig.~\ref{fig:axion_limits} shows the exclusion limits for two scenarios, namely ALPs coupling to 
photons only or also to hypercharge. The exclusion limits presented are the best so far over the m$_{\rm{a}}$ = 5-50 GeV.
\vspace{-1cm} 
\section{Measurement of the inelastic proton-proton cross sections at $\sqrt{s} = 13$ TeV}
\begin{wrapfigure}{r}{6.0cm}
\includegraphics[width=0.45\textwidth]{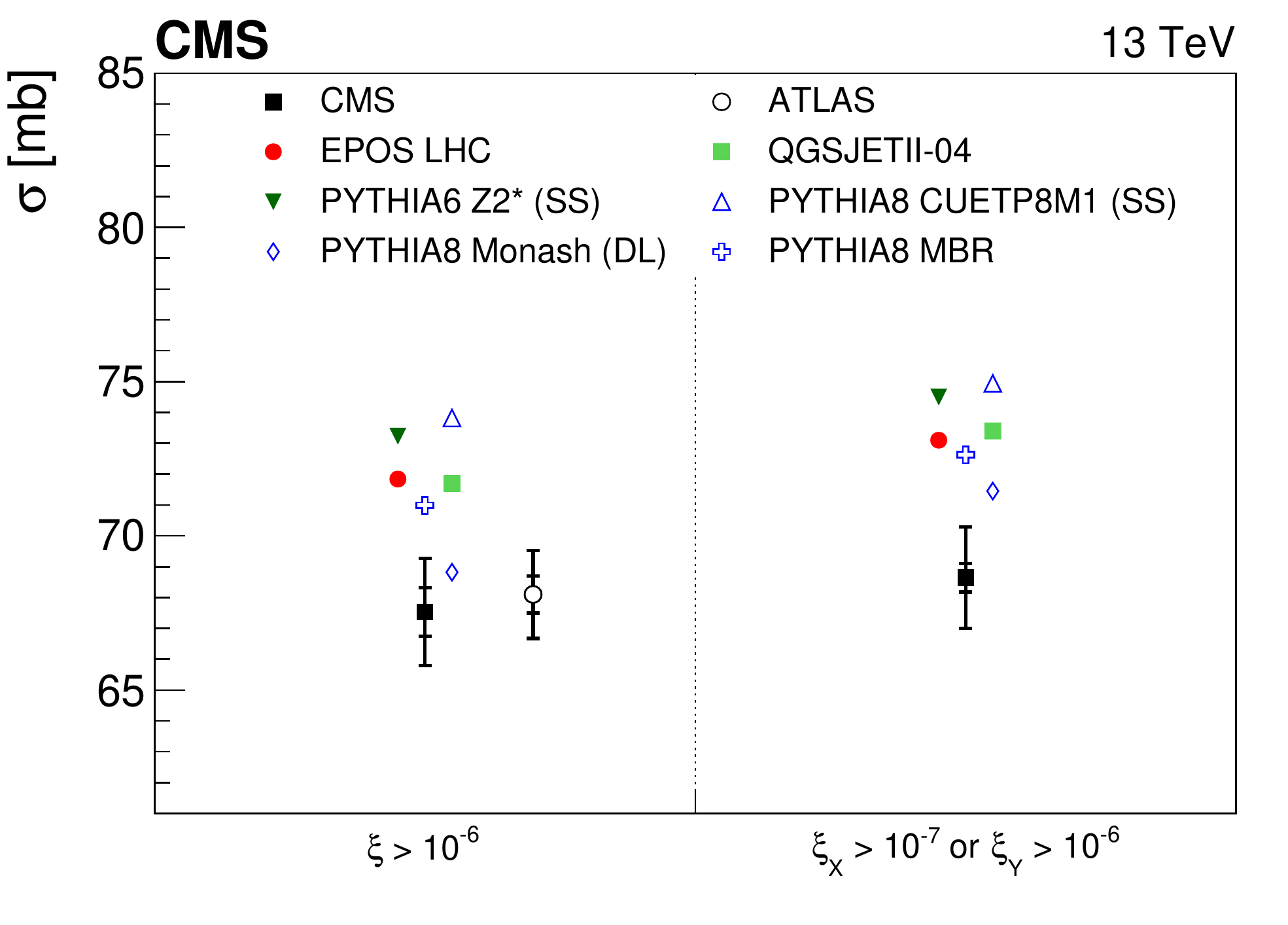}
\caption{Proton-proton inelastic cross section at $\sqrt{s} = 13$ TeV in two phase-space regions \cite{xsec_cms} .} 
\label{fig:inelastic_xsec}
\end{wrapfigure}

A measurement of the inelastic cross section in pp collisions at $\sqrt{s} = 13$ TeV is presented. The analysis includes
the data collected by the HF and CASTOR calorimeter which cover -6.6 $< \eta < $ -3.0 and +3.0 $< \eta < $ 5.2.
This provides a sensitivity to large part of the total inelastic cross section, including diffractive events. 
The fiducial cross section is measured in a phase space region excluding scattered proton fractional momentum
losses at $\xi_{X}=M^{2}_{X}/s < 10^{-7}$ and $\xi_{X}=M^{2}_{Y}/s < 10^{-6}$, corresponding to $M_{X} < 4.1$ GeV
and $M_{Y} < 13$ GeV, where $M_{X}$ and $M_{Y}$ are invariant masses of the dissociated proton systems with negative
and positive pseudorapidities, respectively.  The use of the CASTOR forward calorimeter allows 
the extension of this type of measurement to a low mass region so far unexplored.
Various Monte Carlo event generators are used to correct the measured cross section for
acceptance and instrumental effects, as well as to compare the final results to different
hadron-hadron interaction model predictions. The inelastic events is selected by requiring
an energy deposit above 5 GeV in either of the two HF calorimeters. In the presence of CASTOR, 
inelastic  events  are selected offline by requiring either an energy deposit above 5 GeV in 
either of the two HF calorimeters, or an energy deposit above 5 GeV in CASTOR.
The fully corrected fiducial cross section for HF only calorimeter is 
$\sigma(\xi > 10^{-6}) = 67.5 \pm 0.8 \textrm{(syst)} \pm 1.6 \textrm{(lumi)}$ mb. 
The average cross section obtained from runs with HF and CASTOR calorimeters in 
the extended phase space yields 
$\sigma(\xi_{X} > 10^{-7} \textrm{or}~~\xi_{Y} > 10^{-6}) = 68.6 \pm 0.5 \textrm{(syst)} \pm 1.6\textrm{(lumi)}$ mb ~\cite{xsec_cms}. 
Fig.~\ref{fig:inelastic_xsec} shows the inelastic cross sections in the two phase space domains compared
to the predictions of the various models used in this analysis. 
The measured cross sections are smaller than those predicted by the majority of models.
\vspace{-0.5cm} 
\section{Study of jet gap jet events in pp collisions at $\sqrt{s} = 7$ TeV}
Dijet production in proton-proton collision at the LHC is well described by perturbative 
Quantum Chromodynamics (pQCD) calculations based on DGLAP evolution equations. However, 
when the two jets are separated by a large pseudorapidity interval, a color singlet
exchange (CSE) takes place between the interacting partons which is expected to 
describe the data better by BFKL equations. This report presents the measurement 
of the CSE fraction using data collected by the CMS experiment at $\sqrt{s} = 7$ TeV with low pileup, 
corresponding to an integrated luminosity of 8 pb$^{-1}$. 
Three non-overlapping samples of dijet 
events are used, corresponding to the following three second leading jet p$_{\rm T}$ ranges : 
40-60, 60-100, and 100-200 GeV. In order to allow for a sufficiently wide rapidity gap 
between the jets, the two leading jets are required to be in the range 1.5 $< |\eta| < $ 4.7 
and in opposite hemispheres of the CMS detector. The CSE and non-CSE events are discriminated 
by the charged-particle multiplicity (Ntracks with p$_{\rm T} > 0.2$ GeV) in the gap region between 
the two leading jets. The CSE fraction is defined as the ratio between the number of events 
with gap divided by the total number of dijet events.
\begin{figure*}[htbp]
\begin{center}
\includegraphics[width=0.40\textwidth]{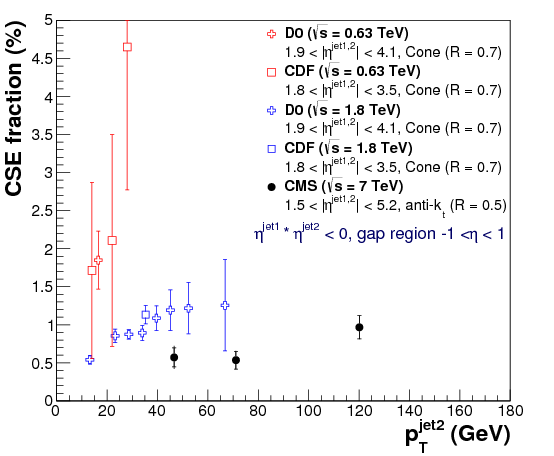}
\includegraphics[width=0.40\textwidth]{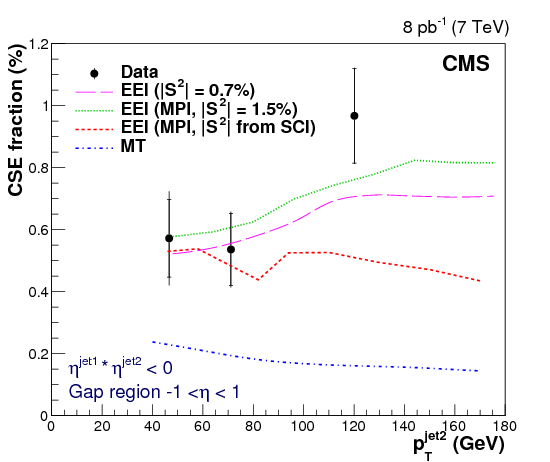}
\caption{CSE fraction as a function of p$_{\rm T}^\textrm{jet2}$ at $\sqrt{s} = 7$ TeV \cite{jpj_cms} 
(a) compared to earlier results, and (b) to the predictions of the Mueller and Tang (MT), 
and EEI model.} 
\label{fig:jet_gap_jet}
\end{center}
\end{figure*}
\vspace{-0.5cm} 
Fig.~\ref{fig:jet_gap_jet}(a) shows fraction of dijet events with a central gap as a function 
of p$_{\rm T}^\textrm{jet2}$, compared to earlier results by D0 and CDF. The decrease
of the gap fraction with increasing $\sqrt{s}$ is in agreement with the earlier 
observation by the TEVATRON experiments. It can be
ascribed to a stronger contribution from rescattering processes, in which the interactions between 
spectator partons destroy the rapidity gap. In Fig.~\ref{fig:jet_gap_jet} (b) the data are compared 
to the BFKL-based theoretical calculations of the MT, and EEI model. 
The MT model does not describe the data, as already observed for the
TEVATRON results. However, EEI model describe many features of the
data, but none of the implementations is able to simultaneously describe all the features of the measurement.

\end{document}